\documentclass[aps,twocolumn,groupedaddress]{revtex4}
\usepackage[section]{placeins}        %Keep Figure in Section
\usepackage{float}% If comment this, figure moves to Page 2
\usepackage{ulem}
\usepackage{epsfig}
\usepackage{pdfpages} %
\usepackage{amsmath,amsthm, amssymb, latexsym}
\usepackage{color}
\usepackage{graphicx}
\usepackage{epsfig}
\usepackage[outercaption]{sidecap}%\usepackage{ulem}

\newcommand{\bee}{\begin{equation}}
\newcommand{\ene}{\end{equation}}
\newcommand{\beea}{\begin{eqnarray}}
\newcommand{\enea}{\end{eqnarray}}

\begin{document}
%\preprint{AIP/123-QED}
\title{Spontaneous formation of coherent structures by  intense laser pulse interacting with overdense plasma}% Force line breaks with \\
%\thanks{Footnote.}
%\author{Devshree Mandal$^1$}
% \email{devshreemandal@gmail.com}

\author{ Devshree Mandal$^{1,2}$}

\thanks{devshreemandal@gmail.com}
\author{AyushiVashistha$^{1,2}$}
\author{Amita Das$^{3}$}

\affiliation{$^1$ Institute for Plasma Research, HBNI, Bhat, Gandhinagar - 382428, India }
\affiliation{$^2$ {Homi Bhabha National Institute, Mumbai, 400094 } }

\affiliation{$^3$  Department of Physics, Indian Institute of Technology Delhi,  Hauz Khas, New Delhi - 110016, India }

%\author{Devshree Mandal, Ayushi Vashistha}%
%\affiliation{ 
% Institute For Plasma Research, HBNI, Gandhinagar, Gujarat, India, Pin 382428%\\This line break forced with \textbackslash\textbackslash
%}%
%\affiliation{ 
%  Homi Bhabha National Institute, Mumbai, 400094, India%\\This line break forced with \textbackslash\textbackslash
%}
%%\author{Atul Kumar$^1$}
%% %\homepage{http://www.Second.institution.edu/~Charlie.Author.}
%
%\author{Amita Das}
%%\affiliation{ 
%%$^1$ Institute For Plasma Research, HBNI, Gandhinagar, Gujarat, India, Pin 382428%\\This line break forced with \textbackslash\textbackslash
%%}%
%
% %\homepage{http://www.Second.institution.edu/~Charlie.Author.}
%\affiliation{ 
%Department of Physics, Indian Institute of Technology, Delhi, India %\\This line break forced% with \\
%}%

% %\homepage{http://www.Second.institution.edu/~Charlie.Author.}
%\affiliation{%
%$^3$Tata institute for Fundamental Research, Mumbai, India%\\This line break forced% with \\
%}
%%\date{\today}% It is always \today, today,
%             %  but any date may be explicitly specified

\begin{abstract} 
 
When a  laser field is incident on an overdense plasma it is unable to penetrate inside it. Nevertheless, a part of its energy gets transferred to the electrons through a variety 
of mechanisms (e.g. vacuum and $\vec{J} \times \vec{B}$ heating \cite{brunel,JXB}). 
The dynamics of these energetic electrons inside the 
plasma is  a field of great interest. It is demonstrated here using   2-D PIC (Particle - In - Cell) simulation that when a high intensity laser pulse is incident on an overdense 
target, energetic electrons get generated which spontaneously 
 organize themselves to form  coherent structures. These coherent structures are observed to have 
 similar dynamical traits as displayed by the 
  solutions of the EMHD (Electron Magnetohydrodynamic) model \cite{Isichenko}.  
  This is noteworthy that EMHD is an approximate model and is 
  applicable for the dynamics of  non-relativistic electrons. However, in these  simulations, the electron  
  energy is in the relativistic regime as the laser intensity is significantly high. Thus, it shows that 
  the relativistic dynamics also permits the existence of robust coherent structures. Interesting   kinetic behaviour at particle level is observed in the simulation which shows that  the coherent structures  take  background electrons in its fold and 
   subsequently  emit them at a higher energy.

 % \textcolor{red}{\textbf{abstract has to be written}}
 
%
%Valid PACS numbers may be entered using the \verb+\pacs{#1}+ command.
\end{abstract}

%Use showkeys class option if keyword
\maketitle

%\begin{quotation}
%\verb+quotation+ environment and is formatted as a single paragraph before %the first section heading. 
%(The \verb+quotation+ environment reverts to its usual meaning after the first sectioning command.) 
%Note that numbered references are allowed in the lead paragraph.
%
%The lead paragraph will only be found in an article being prepared for the journal \textit{Chaos}.
%\end{quotation}

\section{\label{sec:level1}Introduction}

Interaction of laser with plasma has always been an alluring topic with rich  underlying physics. Depending on laser parameters, various aspects of laser plasma interaction has been successfully implemented  in numerous applications ranging from medical therapy\cite{ledingham2004high, medical1, medical2, medical3, macchiRMP}, non-linear optics\cite{macchiRMP,optical1,optical2,optical3} to inertial confinement fusion\cite{temporal2002numerical,roth2001fast,kodama}. The immense potential associated with this generates considerable interest in its study and research. 
Such wide variety of applications have come up because the laser radiation can easily get the plasma to respond  in a  nonlinear fashion. There are several interesting non-linear phenomena which bear testimony to this. The excitation of nonlinear wakefield structures, soliton formation, stimulated Raman scattering, laser focusing etc, are some phenomena which have been observed and widely studied. While most of these 
 listed phenomena are for a plasma which is underdense for  the incident laser radiation. The laser in this case 
  continues to propagate inside the  plasma medium and interacts with the bulk plasma medium. In the context of overdense plasma, the laser energy, however,  gets dumped around the critical density layer 
  (for inhomogenous plasma) 
  and/or at the vacuum plasma interface of an overdense plasma.  The laser energy gets partially  absorbed by the plasma electrons. These energetic electrons then interact with the rest of the plasma and elicit its response in various forms.\\

 We demonstrate here, with the help of 2-D Particle - In - Cell (PIC) simulations that the energetic electrons created by the laser organize together with the background plasma electrons to spontaneously form 2-D coherent structures. We also observe that the  coherent structures exhibit similar characteristics traits 
 as that of the   nonlinear coherent solutions permitted by the Electron Magnetohydrodynamics (EMHD) model. 
% This is interesting as the electron energy in simulation is in the relativistic domain and the EMHD 
% model is fluid depiction of non relativistic electron dynamics. This suggests that there exist 
% robust coherent solutions even in the relativistic domain of electron dynamics. 

The Electron Magnetohydrodynamic (EMHD) model essentially describes the dynamics of magnetized electron fluid for which the time scales of interest are fast enough to ignore ion motion \cite{Isichenko,biskamp} and considerably slower to ignore the displacement current contribution. The EMHD model permits certain coherent nonlinear exact solutions of monopolar and dipole vortex forms which are quite robust and stable. While the monopole vortices are stationary, the dipoles propagate along their axis in a plasma with homogeneous density. When the plasma density is inhomogeneous,  both monopoles and dipoles acquire an additional drift velocity and display interesting dynamical traits. The dynamics of these structures have been studied in detail with the help of fluid simulations by Das et al.\citep{dasppcf} in the context of homogenous plasma and by Sharad et al.\cite{doi:10.1063/1.2943693} for inhomogeneous plasma. 
These coherent structures have been perceived to have important implications as they can possibly be utilized for the purpose of transporting energy from laser to overdense regions of plasma. For instance, this could be useful for igniting the compressed fuel in the fast ignition scenario. In fact, it has been shown earlier in the work by Sharad et al.\cite{SharadK.Yadav} that by appropriately tailoring plasma density inhomogeneity, the path of these structures can be guided and its energy can also be anomalously dissipated at a rate much higher than the permissible limit of classical collisional values\cite{yadav2009anomalous}. 
However, EMHD model depicts electron dynamics in the non-relativistic domain. The utilization 
of the robust, stable coherent solutions of EMHD model for energy transport  would be  inefficient as 
the energy content of such structures would be small. The question, therefore, arises whether 
similar robust solutions exist in the relativistic case and whether they could be excited with intense lasers. \\
% The pertinent question that remains to be answered in this regard is whether such coherent structures can be generated by a laser interaction with an overdense plasma medium and the conditions under which it possibly can be formed. 

 We demonstrate such a possibility by carrying out PIC simulations in the framework of OSIRIS4.0\cite{Hemker, Fonseca2002,osiris}. In this context we will like to mention that some earlier studies have also observed the formation of coherent 2-D structures in simulations. Bulanov et al.\cite{nakamura2010high} were the first to show the generation of dipolar structures in the wake of a laser field in an underdense plasma medium.
Several attempts have been made to study the formation and propagation of self-excited EMHD dipoles in plasmas. In a recent study\cite{mima2015}, external magnetic fields have been employed to guide the magnetic field fluctuations that get generated in the counterstreaming beam plasma systems via Weibel and oblique filamentation instability. The formation of  magnetic dipoles have also been observed in wake of near critical density  plasmas to equilibriate the huge electric field due to evacuating electrons and the resultant magnetic pressure around the depletion region\cite{mima2008}. \\

 In this work, we present a comprehensive study of spontaneous formation and propagation of 
 robust magnetic coherent 
 structures  when an  intense laser falls on  overdense plasma medium.  It has been shown that dipoles are robust when the  density profile of plasma is shallow and they are observed to propagate with faster speed 
 for plasma profile with  sharper density gradient. Particle trajectories  have also been shown which show 
  interesting kinetic behaviour. 
 The paper has been organized as follows. Section II discusses simulation details which has been used in this paper. Section III  contains the observations of the formation of coherent magnetic structures of monopole 
 and dipole kind. Their characteristic propagation dynamics have also been   presented in this section. 
 Section IV discusses the behaviour of electron particles as they interact with this structures and get entrained and subsequently get emitted. Section V contains the concluding remarks. 
 
\section{\label{sec:level2}Simulation Details}
\begin{table}

\caption{Simulation parameters: In normalised units and possible values in standard units. }
	\begin{tabular}{|p{2.5cm}||p{2.5cm}||p{2.5cm}|}
		
		\hline
		\textcolor{red}{Parameters}& \textcolor{red}{Normalised Value}& 	\textcolor{red}{Value in standard unit}\\
		
		\hline
		\hline
		\multicolumn{3}{|c|}{\textcolor{blue}{Laser Parameters}} \\
		\hline
		
		Frequency&$ 0.32$&$3.2 \times 10^{14}$Hz\\
		\hline
		Wavelength& $2.17$ &$1\mu m$\\
		\hline
		Intensity&$a_{0} =25$&$ 8 \times10^{20} W/cm^2$ \\
		\hline
		\multicolumn{3}{|c|}{\textcolor{blue}{Plasma Parameters}} \\
		\hline
		Number density($n_0$)&1 &$3.4 \times 10^{20}$ $ cm^{-3}$\\
		\hline
		Electron Plasma frequency ($\omega_{pe}$)&1&$1\times10^{15}$Hz\\
		\hline
		Electron skin depth ($c/\omega_{pe}$)&1& $0.46\mu m$ \\
	\hline
		
	\end{tabular}

\end{table}	
We have employed OSIRIS 4.0 framework to carry out Particle - In - Cell (PIC) simulation 
\cite{Hemker,Fonseca2002,osiris} for our study. 
We chose a 2-D slab geometry with X-Y plane as our
simulation domain. A square box with each side having a length $L= 500d_{e}$ 
 with $25000 \times 25000$ cells is considered,  ($d_e= c/\omega_{pe} $ is skin depth of the plasma). Fig.\ref{schematic} shows the schematic of the simulation geometry. Region I and IV denote  vacuum 
  whereas region II and III contain a plasma slab. Region II has a linearly increasing plasma density profile along $\hat{x}$ which joins with the  uniform plasma density in region III. Ions are kept stationary and provide a neutralising background. A $p -$polarized laser pulse(with electric field lying in the 2-D plane of simulation) is incident normally from left side of the target. The longitudinal profile of laser pulse is a polynomial function with rise and fall time of $35 \omega_{pe}^{-1}$ and flat top for $10 \omega_{pe}^{-1}$. The laser profile 
  along the transverse $\hat{y}$ is Gaussian with full width at half maxima of $30d_e$. The boundary for particles and electromagnetic fields are absorbing along both the directions. Table I presents laser and plasma parameters in normalized units and a possible  value in standard unit.

 \section{\label{sec:level3}Observations }

 It is well known that a non-relativistic laser incident normally can only penetrate upto a few skin depths of an overdense plasma(\textit{i.e.} $\omega_{L}<\omega_{pe}$).
%When a laser pulse starts interacting with plasma which was initially in equilibrium and its frequency is lower than plasma frequency i.e. $\omega_{L}<\omega_{pe}$ laser pulse will be not able to propagate into the plasma, it can penetrate only upto few skin depths of the plasma.
However, in case of intense pulse(\textit{i.e.} $a_0>1$), under the influence of strong electric field of laser pulse electrons gains a directed velocity($\sim c$) in the system. Thus,  pulse is able to propagate deeper than skin depths as effective plasma frequency gets modified due to relativistic effects\cite{KawDawson,Kaw,Davies_2008}. These energetic electrons accelerating into plasma acts as forward current and background plasma inhibits this strong incoming current by generating a return current in response to it. The interaction of these accelerating electrons with background plasma  is highly non-linear. With time, this non-linearity will make its presence known in the form of coherent structures or a turbulent spectrum. \\

\subsection{Coherent  structures for homogeneous  plasma}
We focus our study on formation of   coherent structure for the case when a laser falls 
on a homogenous plasma slab. For these simulations, therefore, the plasma density 
in region II and region III (Fig.\ref{schematic}) are chosen to be identical. 
The number density in  various regions in this case is chosen as ($n_I = n_{IV}=0, n_{II} = 10.0 n_c, n_{III} = 10n_c$). 
We observe formation of magnetic field 
monopole/ dipole structures  in plasma after its  interaction with an intense laser pulse. The formation of these structures can be attributed to local un-compensation of current in the bulk plasma which leads to spontaneous generation of high magnetic field structures. It has been observed that the forward current (due to energetic electrons) and return current(plasma response to energetic electrons) broadly compensate each other \cite{sentoku} in the beginning. The consequent spatial separation of forward and return currents 
through Weibel instability process is  believed to generate magnetic fields.

Our simulation  demonstrates that  magnetic field with opposite polarity gets generated at vacuum-plasma interface (Fig.~\ref{uniBfld}, t=50) and propagates inwards in the bulk region of a homogeneous 
density plasma. Initially the magnetic field structures are observed to have fine scale structures ($t \approx 100 $). Subsequently,  the magnetic field of similar  polarity 
keep coalescing together to    form  bigger structures. These  structures organize ultimately and  form 
  dipolar magnetic field structures  of opposite polarity. Two such dipolar structures can be seen to form 
  clearly in this figure at $t = 1500$. The lobes of these dipoles are, however, found to have  unequal  strength. 
  
  As discussed earlier the  Electron Magnetohydrodynamics (EMHD) fluid model for electron dynamics permits exact solutions of monopolar and dipolar forms. The monopoles are stationary in inhomogeneous 
  density plasma, whereas dipoles with   perfectly balanced lobes are known to translate 
  along their axis. If, however, the lobe of the dipole are  unbalanced they rotate and also translate 
  along their axis as a result they appear to move along a curved path\cite{dasppcf}. The dipoles formed in this simulation have lobes of dissimilar strength and hence are seen to move along a curved path. The dynamical behaviour of the observed dipolar structures, thus, match with the 
  properties of EMHD solutions. However, it needs to be noted that  the electrons involved in forming these dipolar structures have relativistic speed. This can be observed in the plot 
  of Fig.\ref{spectrum} which shows the energy count of electrons after the laser pulse has hit the plasma target.  It can be observed that there is significant 
   electron count having  energy 
  of the order of  MeV (and even one  order higher). This suggests that the magnetic field 
  structures formed in the simulations involve currents of  electrons which are relativistic. 
  The spontaneous formation of such robust structures by laser and their persistence in simulation suggest that 
   there exist  relativistic coherent solutions  which essentially follow similar dynamical 
  characteristics as displayed by the well known non-relativistic EMHD solutions. 

\subsection{Role of inhomogeneous plasma density on coherent solutions}
We now investigate the formation and propagation of coherent structures for the case when the 
plasma density is inhomogenous. 
Region II shown in schematic(Fig.\ref{schematic}) has a    the plasma density profile which rises linearly and saturates in region III . 
%from zero to $10 n_c$ (where $n_c$ is the critical density).
We provide comparison for two cases of density gradients, Case(A) of sharp density gradient with $dn/dx = 0.05$ and Case(B) with a shallow density gradient of $dn/dx = 0.005$. Clearly, since the width of the regions 
remain fixed in case(A) the plasma density in region II and III are higher than that of case (B).  

The snapshots at various times for the sharp and shallow density profile cases have been shown in Fig.(\ref{nonuniBfld}) and Fig.(\ref{shallown}). It can be observed that in both cases there is formation of dipole structures. 
For case (A) the structures are smaller and sharp. The density being higher than the typical skin depth 
scale is smaller compared to case(B).
The dipole is observed to propagate along its axis towards the high density region. 
However, as it propagates towards the higher density the two lobes gets squeezed together. This can be observed by comparing  the 
 snapshots at $t = 600$ and $t = 900$ in Fig.(\ref{nonuniBfld}). This is the characteristic feature of EMHD dipole propagation\cite{doi:10.1063/1.2943693} wherein 
 an additional drift in the presence of density gradient, $\vec{V_n}=(B\hat{z}\times \vec{\nabla}n)/n^2$, is acquired by the magnetic structures in the presence of density inhomogeneity. 
 This drift for the red lobe is towards positive $y$ axis and for the blue lobe it is in the negative $y$ direction. 
 The axis of the dipole, however, is not perfectly parallel to the density gradient along $x$. Thus, the trajectory of the dipole is 
 oblique and this additional drift turns its axis even further. The mechanism has been demonstrated by fluid simulations of non relativistic 
 EMHD equations by Sharad et al\cite{doi:10.1063/1.2943693}. As a result the two lobes of the dipole exchange there position along $y$ ($t = 2000$). 
 The density gradient induced drift thereafter separates the dipole lobes along $y$. When this separation distance becomes larger than  
 the skin depth the individual lobes act like monopole vortices. These monopole vortices continue to move as a result of 
 density gradient related drift and keep getting separated (snapshots at $t = 3000, 3500, 4000$). \\

It is thus clear that despite the fact that these structures contain relativistic electrons the magnetic coherent structures do exist which essentially correspond to rotating currents in the medium due to electrons. The propagation characteristics of the structures is same as their non relativistic counterpart described by the  EMHD fluid model and its generalization for the inhomogeneous density plasma.  We have also shown that these structures can be excited  directly by lasers when they fall on overdense plasma medium.

 \subsection{ Rotating electron currents by $\vec{E} \times \vec{B}$ drift}
 The monopolar magnetic structures in the plasma essentially represents a rotating electron current in the $X-Y$ plane, which  
  produces magnetic field in the positive or negative $z$ direction depending on whether the electron currents rotate clockwise or anticlockwise. 
   When two oppositely rotating electron currents come within a distance of electron skin depth they form dipole structures. The rotating electron motion once created by the laser is sustained by the  $\vec{E} \times \vec{B}$ drift. The electric $\vec{E}$ and the magnetic field $\vec{B}$ in turn are self consistently generated by the electron flow. This has been illustrated clearly in the plots of 
   Fig.(\ref{uniBfldquvier}) and Fig.(\ref{quiver}) by superimposing the quiver plot of the electric field on the surface plot of 
   magnetic field. The $\vec{E} \times \vec{B}$ drift is clearly directed along the   rotational flow of the electron fluid as expected.

  \section{ Individual electron trajectories in dipole structures}
  The electron as a fluid clearly moves in with  $\vec{E} \times \vec{B}$ drift velocity in these structures. However, one would be interested to know how the individual electrons behave in these structures. 
  Do the same set of electron remain in the structure or do new electrons from the background plasma 
keep getting entrained and emitted by the structure as it moves.  How  does the structure interact with 
background  individual electron etc., are 
some  issues which we explore here. Some issues pertaining to 
 particle level description for the monopole and 
dipolar solutions of EMHD model 
have been studied  earlier\cite{hata2013, Hata_2016}. 
  
   We have tracked the dynamics of several electron   particles near but outside the dipole structure. We 
   show in Fig.({\ref{particle}) 
   the trajectory of one such particle and the evolution of its kinetic energy. The particle can be observed to first get entrained inside the dipole structure. In addition of $\vec{E} \times\vec{B}$ rotation  
   in the dipole lobe the particle also exhibits the gyro-motion in the inhomogeneous magnetic field. 
   After remaining inside the structure for a long time the particle can be observed to leave the structure 
   and squirt out with high speed. In the entire duration the kinetic 
   energy of the particle is observed to fluctuate rapidly. 
   It can be observed from the plot that 
   there are occasions when the fluctuation amplitude of the particle is suddenly quite high.  
   We have marked these time by $t_1, t_2, t_3, t_4$ and $t_5$. We observe that these events occur when the particle trajectory jumps from one lobe to other. In Fig.\ref{crossover}, we have shown the location of the particle before and after these events and it can be clearly seen that the particle has jumped from one location to another 
   when such events occur. This can be understood by realising that as the particle crosses over from one 
   lobe of the dipole to other it passes through a magnetic null. This is similar to 
    the magnetic reconnection geometry where  electron acceleration occurs. 
    It has also been observed that the electrons after getting trapped in the structure ultimately get 
ejected from it with  high kinetic energy. Thereafter,  the particle moves uninterrupted in the plasma. 
One can thus expect the possibility of energetic electrons preceding the arrival of these structures 
in deeper region of the plasma target.

 \section{\label{sec:level4}Conclusions}
  We simulated interaction of an intense laser pulse with overdense plasma and observed spontaneous formation of 
  coherent magentic  structures in the system. These structures are found to be fairly robust and follow the dynamics displayed by 
  solutions of the EMHD fluid model despite the fact that the electrons involved in the formation of these structures have relativistic energies. 
  Thus we have demonstrated the existence of 2-D coherent magnetic structures for relativistic electron fluid. These structures can move in the overdense region of the plasma and hence can be ideal for energy transport. 
  We have also shown that   such coherent structures can be excited by lasers directly as it interacts with the overdense  plasma target. 
  
The particle level kinetics is observed to be interesting for these structures. 
The background low energy electrons are observed to  get trapped in these structures. Some spend a considerable time in these structures. 
These electrons are seen to rotate with the structure and also gyrate in the inhomogenous magnetic field of the dipole. The kinetic energy of the particles show rapid oscillations as the particles rotate in the magnetic lobes. It is observed that a huge jump in the kinetic energy is accompanied 
as the particles jump from one lobe of the dipole to other. It is normally observed that the electrons ejected out of the structures have a 
much higher kinetic energy than the background electron particles and they squirt out of the structure with very high directed velocity 
moving inside the deeper regions of the plasma target.

 \section*{Acknowledgements}
 The authors would like to acknowledge the OSIRIS Consortium, consisting of UCLA and IST(Lisbon, Portugal) for providing access to the OSIRIS4.0 framework which is the work supported by NSF ACI-1339893. AD
would like to acknowledge her J. C. Bose fellowship
grant JCB/2017/000055 and the CRG/2018/000624
grant of DST for the work. The simulations for the work described in this paper were performed on Uday, an IPR Linux cluster. DM  and AV would like to thank Mr. Omstavan Samant for fruitful discussions at IPR.

%
%lon_type = "polynomial",  

\paragraph*{\bf{References:}}
\bibliographystyle{unsrt}
%\figurestyle{unsrt}

\bibliography{ref_dev}

	\begin{widetext}
	\begin{figure*}
\centering
\includegraphics[width=\textwidth]{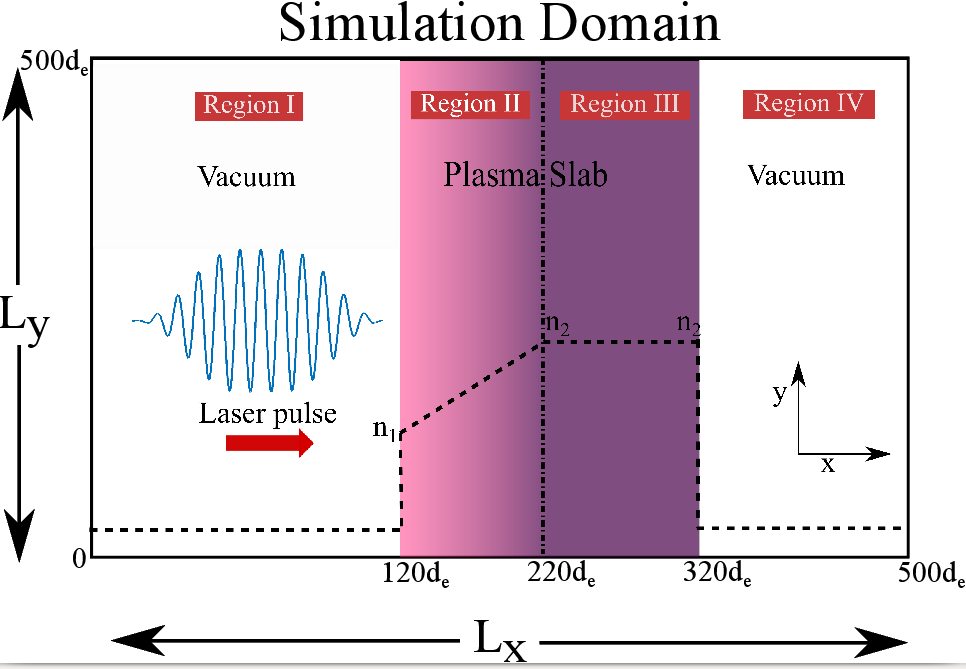}
\caption{Schematic of simulation setup(not to scale)used for this study. Simulation plane has been divided into four regions where Region I and IV are vacuum and region II and III contains plasma. A finite longitudinal and transverse extent intense pulse is incident on this plasma slab. }
\label{schematic}
\end{figure*}

\begin{figure}
	\centering
		\includegraphics[width=\textwidth]{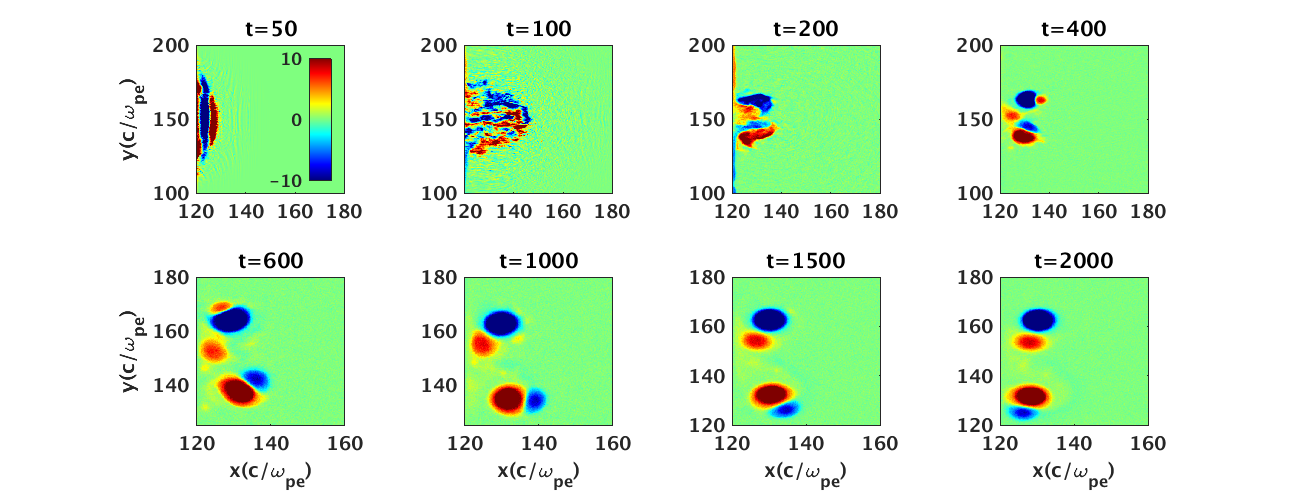}
		\caption[width=1.0\textwidth]{Evolution of $B_z$ where plasma slab has homogenous density profile i.e. $n=10n_c$ from $x=120 $ to $x=320$.  } \label{uniBfld}
	
\end{figure}

\begin{figure}
	\centering
		\includegraphics[width=\textwidth]{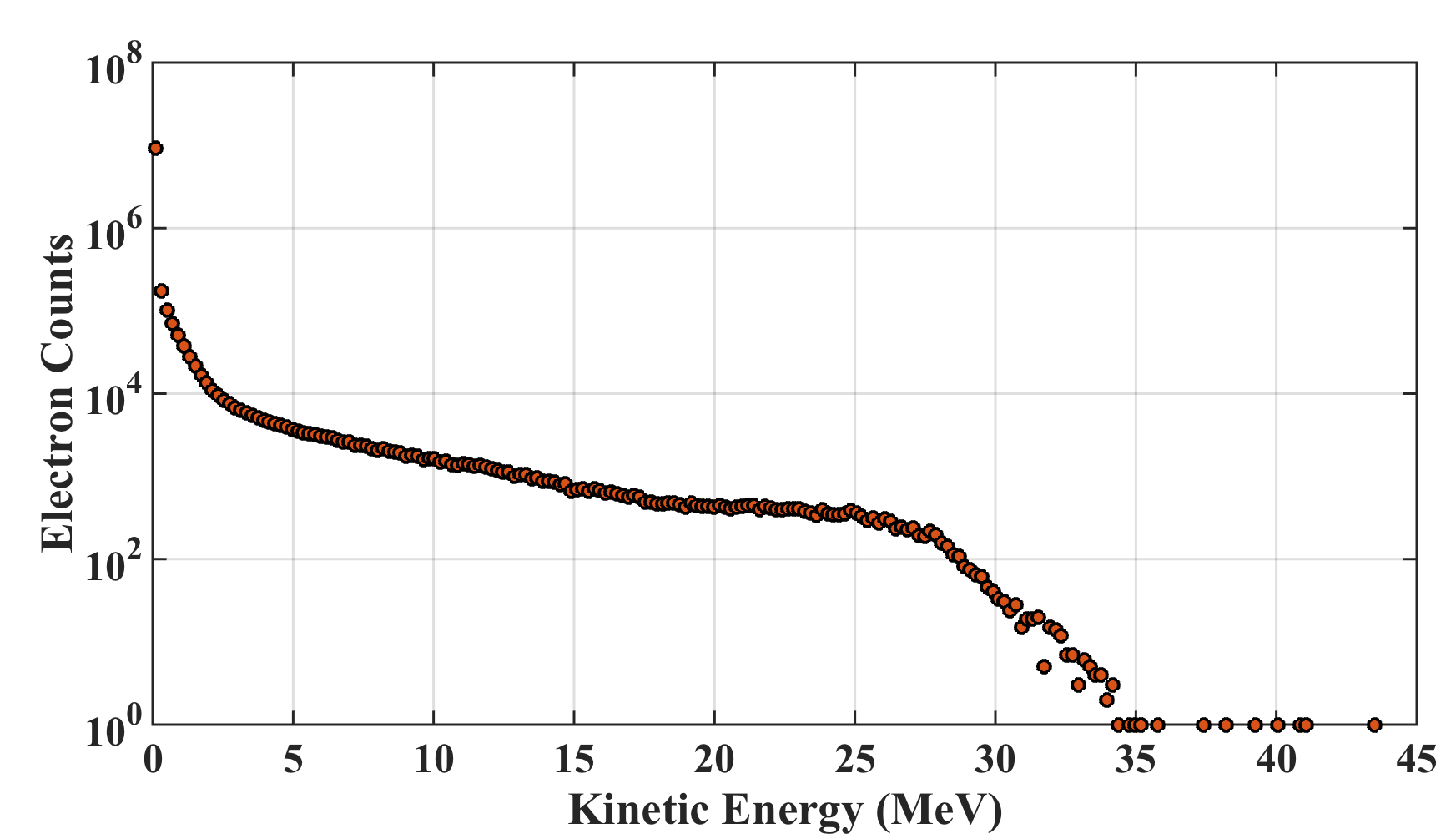}
		\caption[width=1.0\textwidth]{ Electron energy spectrum at t=450, as can be observe electron energy goes upto 30-40 MeV estabilishing the relativistic nature of dipole and particles in it. } \label{spectrum}
	
\end{figure}

\begin{figure}
\centering
\includegraphics[width=\textwidth]{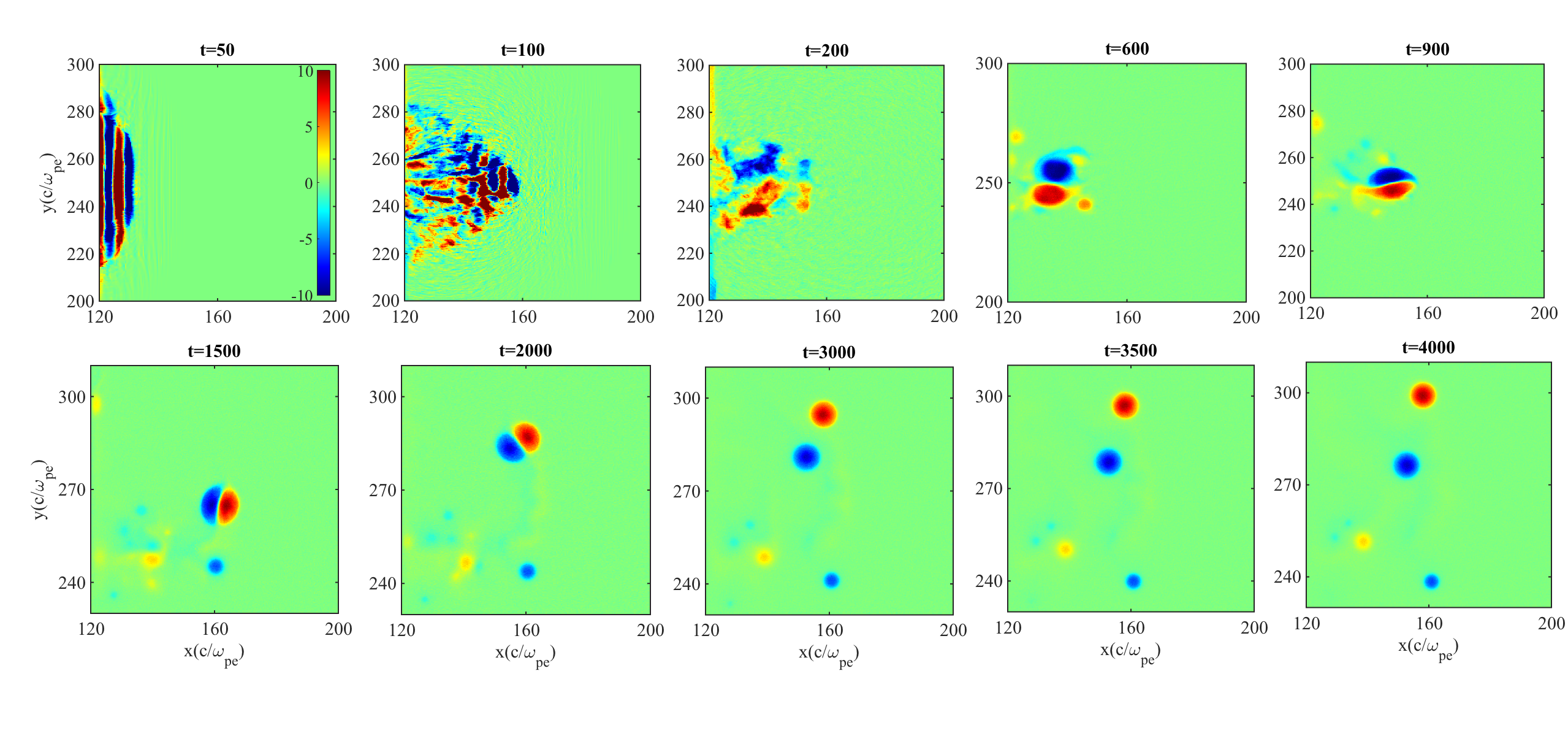}
 \caption{Evolution of $B_z$ where plasma slab has inhomogenous density profile i.e. $\delta n/\delta x= 0.05 $ in region II along $\hat{x}$. }
 \label{nonuniBfld}
\end{figure}

\begin{figure}
\centering
\includegraphics[width=\textwidth]{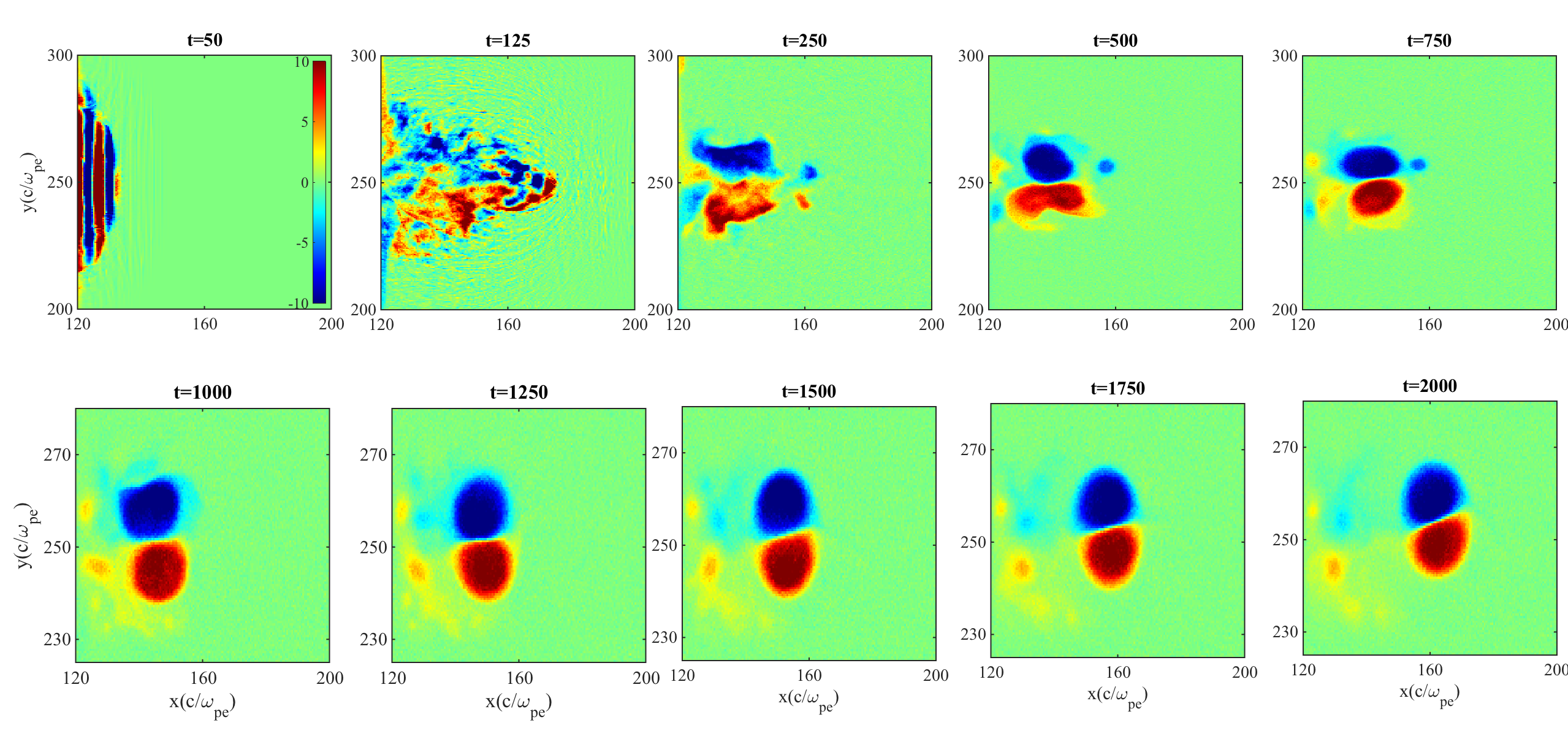}
 \caption{    Evolution of $B_z$ where plasma slab has inhomogenous density profile i.e. $\delta n/\delta x= 0.005 $ in region II along $\hat{x}$.  }
\label{shallown}
\end{figure}

\begin{figure}
	\centering
		\includegraphics[width=\textwidth]{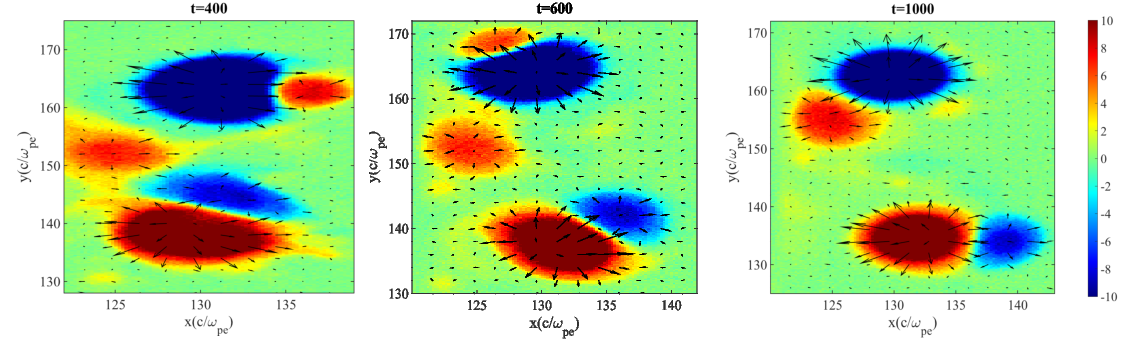}
		\caption[width=1.0\textwidth]{Quiver plot of $\vec{E}$ is being plotted over colorplot of $B_z$  to show the effect of $\vec{E} \times \vec{B}$ in the homogenous case where $n=10n_c$. } \label{uniBfldquvier}
	
\end{figure}
\begin{figure}
\centering
	\includegraphics[width=\textwidth]{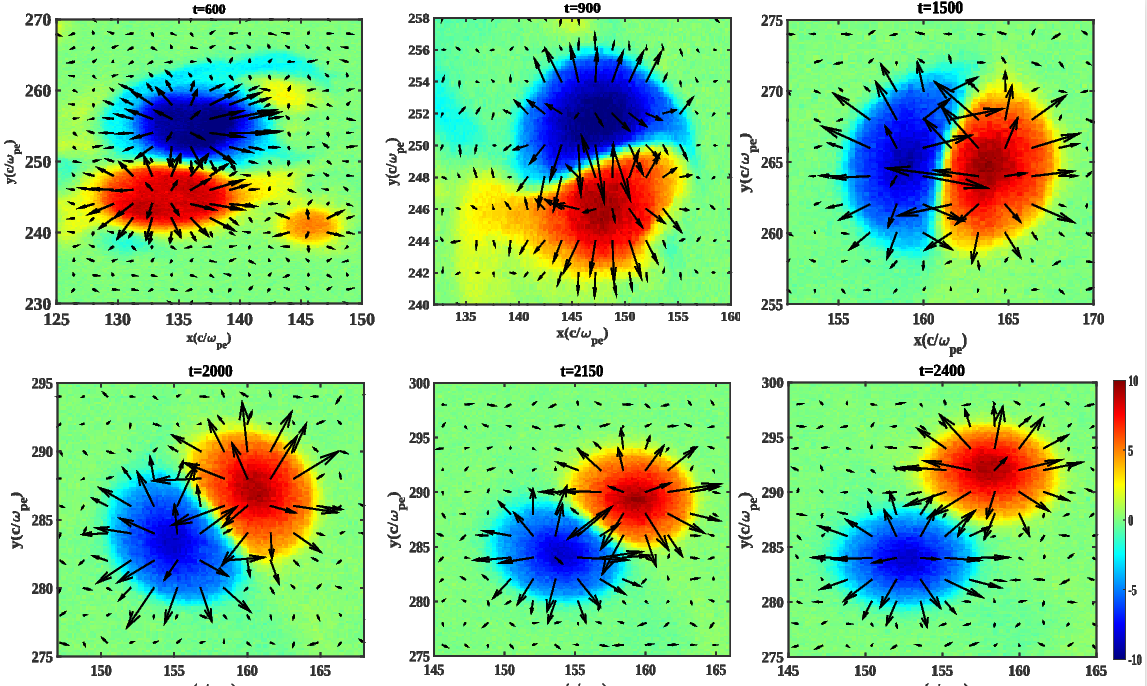}
	\caption{ Quiver plot of $\vec{E} $ is being plotted over colorplot of $B_z$  to show the effect of $\vec{E} \times \vec{B}$ in the inhomogenous case where $\delta n / \delta x = 0.05$.  }
	\label{quiver}
\end{figure} 

\begin{figure}
\centering

	\includegraphics[width=\textwidth]{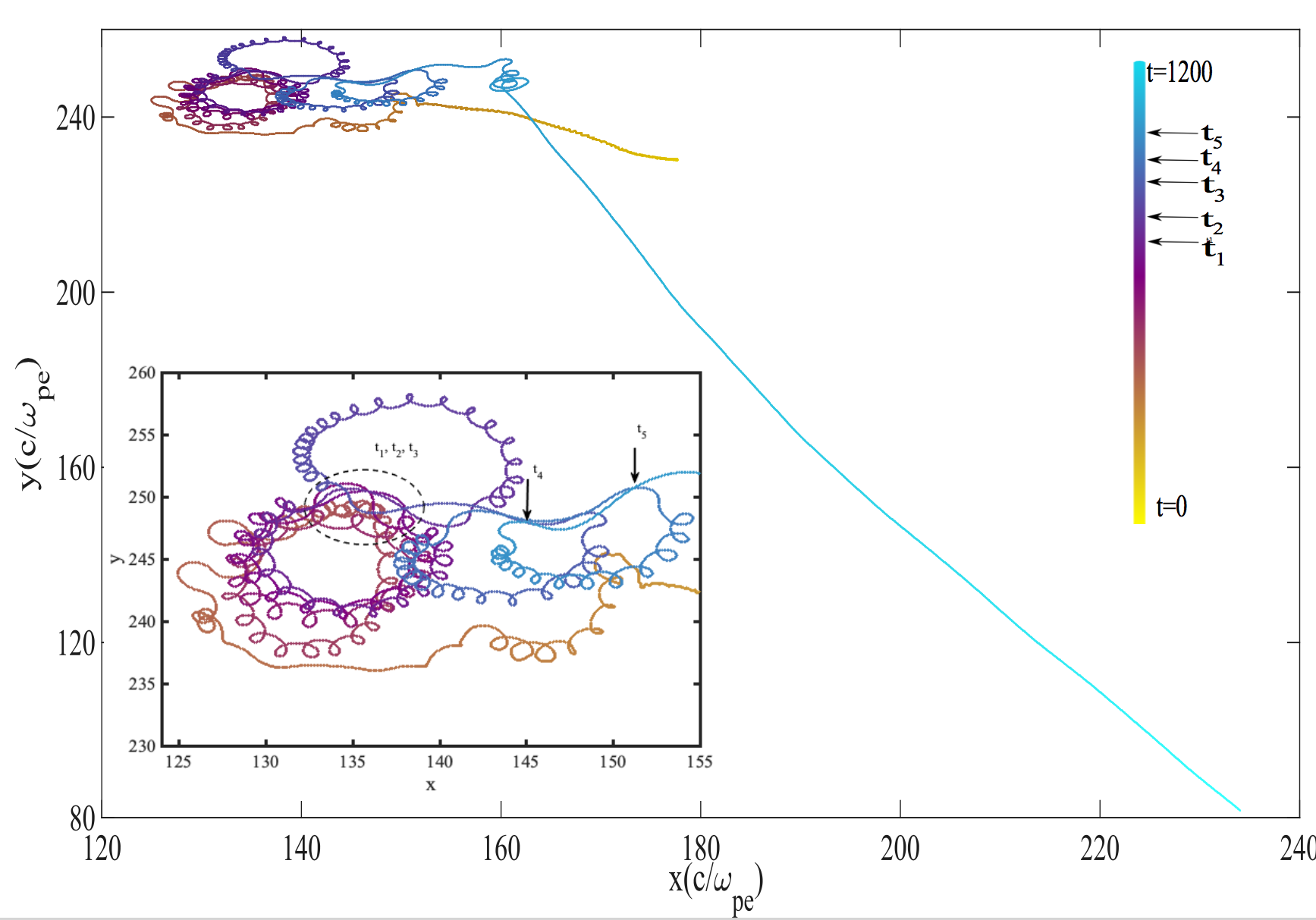}
	\textbf{(a) particle trajectory}\par\medskip
	\includegraphics[width=1\textwidth]{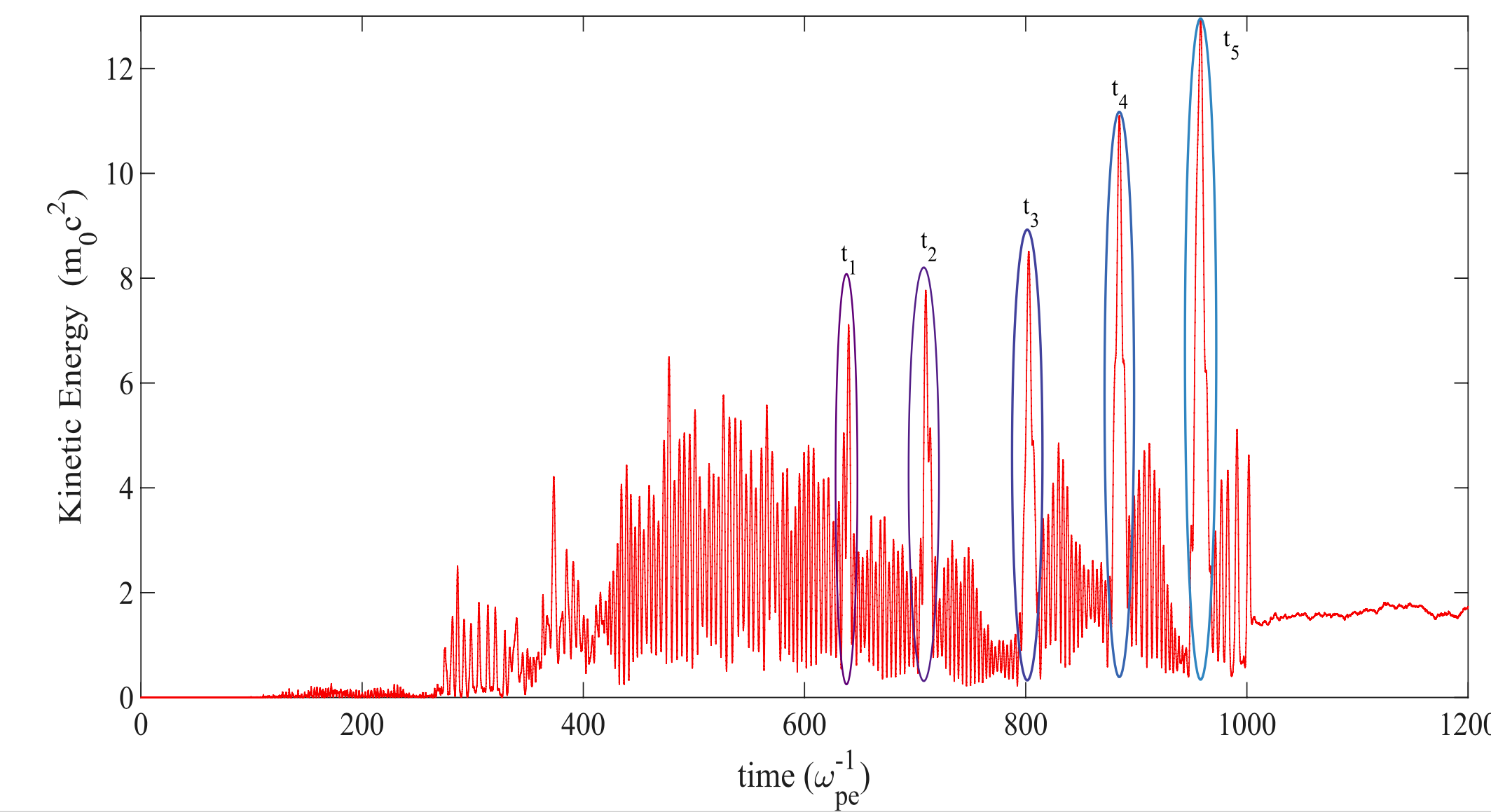}
	\textbf{(b)Kinetic Energy of particle Vs time}\par\medskip
	\caption{At initial time, particle is entrailed by dipole fields and at later time it is ejected from the dipole with a constant energy. At $t_1$,$t_2$, $t_3$,$t_4$ and $t_5$  are those time snapshots when  crossover of electron happens from one lobe of dipole to other lobe. This exchange of dipole results in kinetic energy peak of electron evidently shown in subplot(b) }
	\label{particle}
\end{figure} 

\begin{figure}
\centering
	\includegraphics[width=\textwidth]{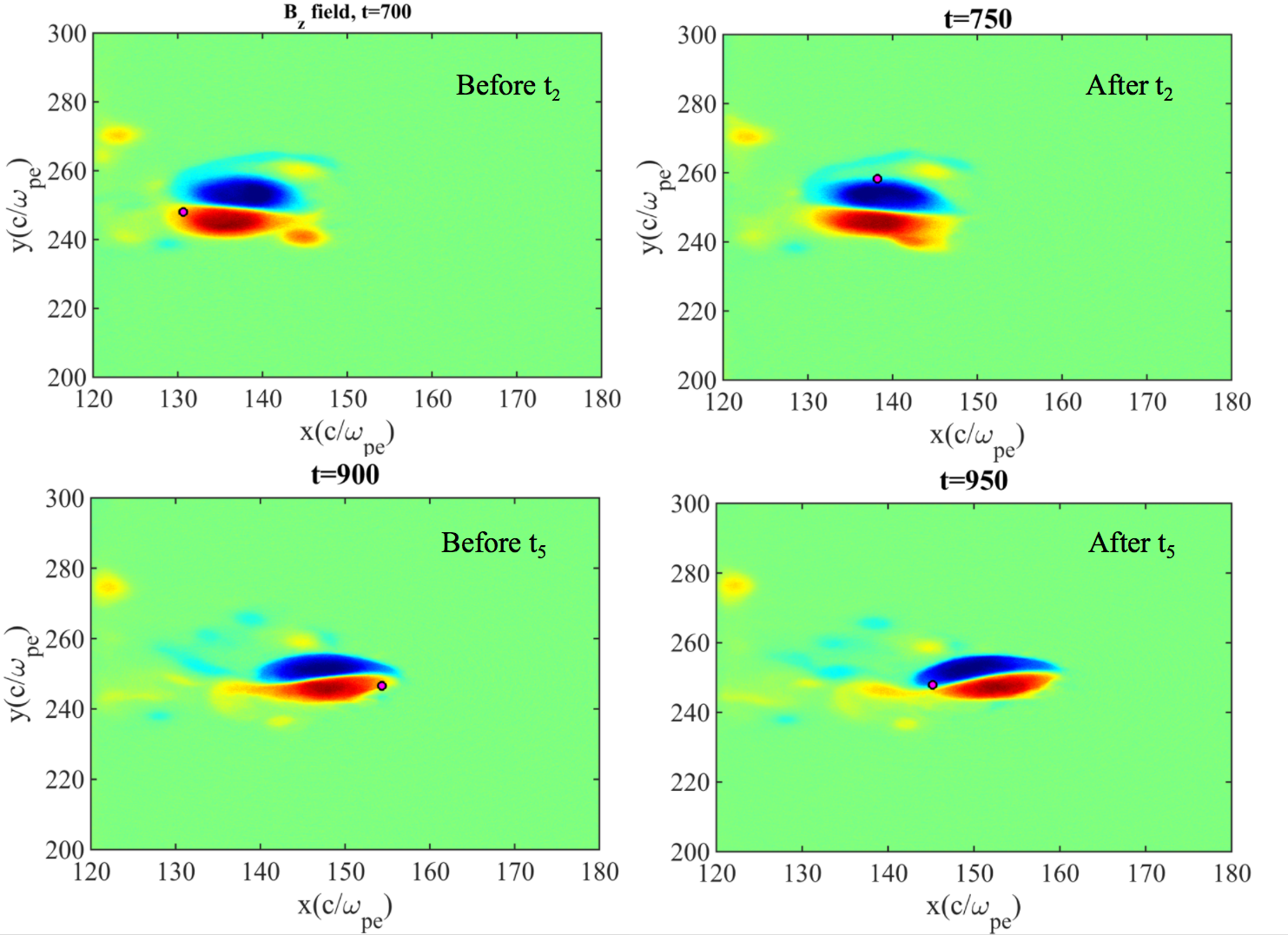}
	\caption{   Particle (magenta color) cross over from one lobe of dipole to another and while doing so its kinetic energy shoots up. We have shown here particle position in dipole before and after such jumps of kinetic energy. }
	\label{crossover}
\end{figure} 

\end{widetext}
\end{document}